\documentclass[aps,12pt,showpacs]{revtex4}%
\usepackage{amsfonts}
\usepackage{amsmath}
\usepackage{amssymb}
\usepackage{graphicx}%
\providecommand{\U}[1]{\protect\rule{.1in}{.1in}}
\providecommand{\U}[1]{\protect\rule{.1in}{.1in}}

\begin{document}
\title{Collapsing Layers on Schwarzschild-Lemaitre Geodesics}
\author{J.P. Krisch and E.N. Glass}
\affiliation{Department of Physics, University of Michigan, Ann Arbor, MI 48109}
\date{5 September 2007}

\begin{abstract}
We discuss Israel layers collapsing inward from rest at infinity along
Schwarzschild-Lemaitre geodesics. The dynamics of the collapsing layer and its
equation of state are developed. There is a general equation of state which is
approximately polytropic in the limit of very low pressure.\ The equation of
state establishes a new limit on the stress-density ratio.

\end{abstract}

\maketitle

\section{Introduction}

Since their introduction, Israel layers \cite{Isr66},\cite{Isr67},\cite{Bar91}
have played an increasingly important role in gravitational physics. Barrabes
and Israel \cite{Bar91} began their paper with a description of the Israel
layer as a thermodynamic phase boundary, but the initial applications of
Israel layers considered metric matching in dynamic collapse processes
involving dust shells, null shells, and cosmic string loops \cite{Isr66}%
,\cite{Isr67},\cite{Bar91}. Poisson \cite{Poi04} has summarized some of the
early seminal work by Israel \cite{Isr66},\cite{Isr67}, Barrabes \cite{Bar91},
de la Cruz \cite{CI68}, Musgrave and Lake \cite{ML97}, and Barrabes and Hogan
\cite{BH98}. The Israel junction conditions \cite{Isr66},\cite{Isr67} find
wide application because they provide a simple dynamic boundary description
for a variety of scenarios \cite{BKT87}. The use of layers as boundaries
between standard metrics has been summarized by Ansoldi \cite{Ans02},
and\ there is a growing literature that applies the thin shell formalism to
new areas such as shell quantum mechanics \cite{Ans02} - \cite{AJ05}, layer
dimensionality \cite{Kri05},\cite{BGK06},\cite{Kri06}, layers in extensions of
general relativity \cite{GW07} - \cite{BI05} and matching in perturbed
spacetimes \cite{LC05} - \cite{MMV07}. Astrophysical problems using layers
with constant spacetime character continue to be interesting, and include the
investigation of collapse \cite{KMM06},\cite{GV89},\cite{LL07} and new
phenomena such as gravastars \cite{MM01} - \cite{BN07}.\ A review of recent
layer applications reveals increasingly complex layer models, the gravastar
models, for example, involving multi-layer constructions, and there is an
increasing interest in describing more realistic layers \cite{KM02}%
,\cite{KKM06}.

A layer metric is determined by the metrics of the two bounding
manifolds.\ The stress-energy of the layer is determined by jumps in the
extrinsic curvatures of the metrics on either side of the layer. For example,
the Israel layer between exterior vacuum Schwarzschild and interior Minkowski
has a 2+1 metric as seen from both sides of the layer \cite{Poi04}
\begin{equation}
ds^{2}=-d\tau^{2}+R^{2}(\tau)d\Omega^{2}. \label{metric-1}%
\end{equation}
The development of $R(\tau)$ tracks the dynamics of the layer and, as in any
general relativity solution with stress-energy, the equation of state of the
layer is important in characterizing the dynamics.

In this work we discuss layers which start from rest at infinity in an
exterior Schwarzschild metric with mass parameter $m_{0}$ and drop inward
along a geodesic.\ The Lemaitre form of the Schwarzschild metric is adapted to
this particular geodesic motion but the layer motion can be simply described
with the usual Schwarzschild metric.\ The interior space is Schwarzschild with
mass parameter $M$.\ Although the motion of the layer is simple, the equation
of state is cubic in the stress, quartic in the density and it provides a new
restriction on the range of the stress/density ratio for collapsing layers.\ 

Layers with simple motions, such as the layer considered here, can require
more physical content than the density and pressure coming from simple perfect
fluid or polytropic models. We quote from \cite{KMM06}: "... models satisfying
a closed equation of state are relatively few and essentially restricted to
dust or linear barotropic models. However, it is worth stressing the
importance of considering general equations of state in realistic models,
describing for instance, very high density regimes for white dwarfs and
neutron stars." Layers with physically realistic equations of state are
important in studying the collapse scenarios that result in objects like
neutron stars. \ 

In the next section we describe the time development of the radial function,
$R(\tau),$ in the exterior Schwarzschild metric. The stress-energy, the
equation of state of the layer and its symmetries are discussed in Section
III, and we close with a Discussion. Extrinsic curvatures are computed in an Appendix.

\section{Layer Geometry}

\subsection{Bounding metrics}

The 2+1 layer is bounded by two metrics%
\begin{equation}
ds^{2}=g_{ab\pm}dx^{a}dx^{b}=-f_{\pm}dt^{2}+(1/f_{\pm})dr^{2}+r^{2}d\Omega^{2}
\label{2-metrics}%
\end{equation}
with +/- denoting exterior/interior.
\begin{subequations}
\begin{align}
f_{+}  &  =1-2m_{0}/r\\
f_{-}  &  =1-2M/r.
\end{align}
The metrics correspond to an $m_{0}$ Schwarschild exterior and an $M=const$
Schwarzschild interior. The layer is described by $r=R(\tau)$ and $t=T_{\pm
}(\tau)$. The velocity of the layer as seen by observers moving with the layer
in the bounding spaces is $U_{\pm}^{a}=(\dot{T}_{\pm},\dot{R},0,0)$, with
corresponding normal vector $n_{a\pm}=(-\dot{R},\dot{T}_{\pm},0,0)$
\cite{Poi04}.\ The normal vector is chosen to be outward pointing, toward the
exterior spacetime. The velocity normalization imposes the condition%
\end{subequations}
\begin{equation}
f_{\pm}(R)\dot{T}_{\pm}^{2}-\frac{\dot{R}^{2}}{f_{\pm}(R)}=1.
\end{equation}
This normalization is used to define function $\beta_{\pm}$%
\begin{equation}
\beta_{\pm}^{2}:=f_{\pm}^{2}(R)\dot{T}_{\pm}^{2}=f_{\pm}(R)+\dot{R}%
^{2}\text{.}%
\end{equation}
$\beta_{\pm}$ is chosen positive with $\dot{T}_{\pm}>0$ and $\dot{R}<0$
describing an infalling layer. \ An expanding layer can be described by proper
time inversion. \ The equation of state of the layer will not depend on the
direction of motion. The induced metric on the layer is%
\begin{equation}
h_{ij\pm}:=\frac{\partial x^{a}}{\partial x^{i}}\frac{\partial x^{b}}{\partial
x^{j}}g_{ab\pm}%
\end{equation}
with%
\begin{equation}
h_{\tau\tau\pm}=-(\frac{\partial T_{\pm}}{\partial\tau})^{2}f_{\pm}%
+(\frac{\partial R}{\partial\tau})^{2}\frac{1}{f_{\pm}}.
\end{equation}
$h_{\tau\tau\pm}=-1$ from the velocity normalization. The metric of the layer
from both sides of the boundary is given by Eq.(\ref{metric-1}).

In the next section, $R(\tau)$ will be fixed by the requirement of geodesic
layer motion in the exterior spacetime. Since both sides of the layer will
agree on the layer metric,\ this will determine the interior radial function.

\subsection{Exterior geodesics}

For Schwarzschild geodesics in the exterior spacetime we have%
\begin{align}
\frac{dR}{d\tau}  &  =\pm\sqrt{E_{0}^{2}-1+2m_{0}/R}\\
\frac{dT_{+}}{d\tau}  &  =\frac{E_{0}}{1-2m_{0}/R}%
\end{align}
where $E_{0}$ describes the initial point of the layer motion.\ $E_{0}=1$,
$dR/d\tau<1$ corresponds to a layer beginning its inward drop from infinity.
The relations to use in describing the infalling layer from the exterior
spacetime are%
\begin{align}
\frac{dR}{d\tau}  &  =-\sqrt{2m_{0}/R}\\
\frac{dT_{+}}{d\tau}  &  =\frac{1}{1-2m_{0}/R}.
\end{align}
Integrating, one finds%
\begin{equation}
R(\tau)=(2m_{0})^{1/3}[(3/2)(c_{1}-\tau)]^{2/3}\text{,} \label{cap-R}%
\end{equation}
which completely determines the layer dynamics. The metric of the layer is%
\begin{equation}
ds^{2}=-d\tau^{2}+[(3/2)\sqrt{2m_{0}}]^{4/3}(c_{1}-\tau)^{4/3}d\Omega^{2}.
\end{equation}

\section{Layer Stress-Energy}

\subsection{Fluid description}

We assume a perfect fluid stress-energy for the layer. The co-moving velocity
is $U^{i}=(1,0,0)$ with density $\sigma$ and stress-energy
\begin{equation}
S_{\ j}^{i}:=(\sigma+P)U^{i}U_{j}+Ph_{\ j}^{i}.
\end{equation}
The stress-energy of the 2+1 layer is related to jumps in the extrinsic
curvatures \cite{Isr66},\cite{Isr67} of the bounding metrics
\begin{subequations}
\begin{align}
-8\pi S_{\ j}^{i}  &  :=\ <K_{j}^{i}>-<K>h_{j}^{i}\nonumber\\
-8\pi S_{\tau}^{\tau}  &  =-<K_{\theta}^{\theta}+K_{\phi}^{\phi}>\\
-8\pi S_{\theta}^{\theta}  &  =-<K_{\phi}^{\phi}+K_{\tau}^{\tau}>\\
-8\pi S_{\phi}^{\phi}  &  =-<K_{\theta}^{\theta}+K_{\tau}^{\tau}>,
\end{align}
where $h_{ij}=[-1,R^{2},R^{2}\sin^{2}\theta]$, $h_{j}^{i}=\delta_{j}^{i}$ and
where
\end{subequations}
\begin{equation}
<K_{j}^{i}>\ :=K_{j+}^{i}-K_{j-}^{i}%
\end{equation}
is the jump in the extrinsic curvature and $K:=K_{i}^{i}$. The extrinsic
curvatures are worked\ out in Appendix A. For the constant mass exterior
Schwarzschild metric they are%
\begin{align}
K_{\theta+}^{\theta}  &  =K_{\phi+}^{\phi}=1/R\\
K_{\tau+}^{\tau}  &  =0.
\end{align}
In the interior, with mass parameter $M$, the extrinsic curvatures are%
\begin{align}
K_{\theta-}^{\theta}  &  =K_{\phi-}^{\phi}=\beta_{-}/R\\
K_{\tau-}^{\tau}  &  =\frac{M-m_{0}}{R^{2}\beta_{-}}.
\end{align}
The layer density is
\begin{equation}
4\pi\sigma=-<K_{\theta}^{\theta}>\ =\frac{\beta_{-}-1}{R} \label{sigma}%
\end{equation}
where%
\begin{equation}
\beta_{-}=\sqrt{1+2\frac{m_{0}-M}{R}}. \label{beta-minus}%
\end{equation}
For $R>0$, the range of $\beta_{-}$ is restricted to $\beta_{-}>1$ for
positive layer density. This requires $m_{0}>M$. The stress is%
\begin{align}
8\pi P  &  =\ <K_{\theta}^{\theta}+K_{\tau}^{\tau}>\text{ }=\frac{-\beta
_{-}+1}{R}+\frac{m_{0}-M}{R^{2}\beta_{-}}\nonumber\\
8\pi P  &  =-\frac{(\beta_{-}-1)^{2}}{2\beta_{-}}\frac{1}{R}. \label{pressure}%
\end{align}
A layer mass $m_{L}$ can be calculated from the layer density
\begin{align}
m_{L}  &  =4\pi\sigma R^{2}\nonumber\\
m_{L}  &  =\frac{2(m_{0}-M)}{\beta_{-}+1}. \label{m-L}%
\end{align}
The physical picture has an exterior Schwarzschild metric with mass parameter
$m_{0}$ formed from the stress-energy of the layer plus the interior
Schwarzschild metric with mass $M$. The binding energy of the layer
contributes to the value of $m_{0}$, as can be seen by eliminating $\beta_{-}$
between equations (\ref{beta-minus}) and (\ref{m-L}):
\begin{equation}
m_{L}=m_{0}-M-\frac{m_{L}^{2}}{2R}.
\end{equation}
The layer mass is not constant due to work done by the stress%
\begin{equation}
\frac{d(\sigma R^{2})}{d\tau}=-P\frac{dR^{2}}{d\tau}.
\end{equation}
This has also been noted by Carr and Yahil \cite{CY90} for a self-similar
Friedman universe and by Visser \cite{Vis95} in dynamic wormholes.

\subsection{Equation of State}

The relation between the density and stress for Lemaitre layers is%
\begin{equation}
P=-\sigma\frac{\beta_{-}-1}{4\beta_{-}} \label{p-sigma}%
\end{equation}
or%
\begin{equation}
\beta_{-}=\frac{1}{1+4P/\sigma}. \label{beta-p-sig}%
\end{equation}
Expression (\ref{beta-p-sig}) can be rewritten as%
\[
\beta_{-}-1=-\frac{4P/\sigma}{1+4P/\sigma},\text{ \ }\beta_{-}+1=\frac
{2+4P/\sigma}{1+4P/\sigma}%
\]
or%
\[
\beta_{-}^{2}-1=-8(P/\sigma)\frac{1+2P/\sigma}{(1+4P/\sigma)^{2}}.
\]
Eliminating $R$ between Equations (\ref{sigma}) and (\ref{beta-minus}) allows
$\sigma$ to be expressed as%
\[
\sigma=\frac{(\beta_{-}-1)(\beta_{-}^{2}-1)}{8\pi(m_{0}-M)}.
\]
Substituting for $(\beta_{-}-1)$ and $(\beta_{-}^{2}-1)$ in terms of
$P/\sigma$ above, yields the equation of state for the Lemaitre layer
\begin{equation}
P^{2}=[\frac{\pi(m_{0}-M)}{4}]\frac{\sigma^{3}(1+4P/\sigma)^{3}}%
{(1+2P/\sigma)}. \label{p-square}%
\end{equation}
This expression can be rewritten for the density or stress in terms of the
ratio $x:=P/\sigma.$%
\begin{equation}
\sigma=\frac{4x^{2}(1+2x)}{\pi(m_{0}-M)(1+4x)^{3}}.
\end{equation}
For a static membrane under tension, $x$ is the negative of the sound speed
squared, $\partial P/\partial\sigma$. The '$x$' ranges of possible interest
are
\begin{subequations}
\begin{align}
I  &  :0\leq x\label{x-ranges}\\
II  &  :-1/4\leq x\leq0\\
III  &  :-1/2<x\leq-1/4\\
IV  &  :-1\leq x<-1/2
\end{align}
The relative size of $m_{0}$ to $M$ provides three mass regions to consider.
$m_{0}=M$ has no extrinsic curvature jumps and no layer.$\ $For $m_{0}>M$,
$\beta_{-}>1$ is required for positive layer density.\ This eliminates the
positive range of $x.$ $\beta_{-}>0$ requires $-1/4\leq x$ and this eliminates
the third and fourth region. Region II is the only allowed range and this sets
a new limit on the stress/density ratio for layers collapsing in a
Schwarzschild geometry. Figure 1 shows the variation of the stress and density
in Region II.\ A portion of Region I has been included for clarity.%
%TCIMACRO{\FRAME{fhFU}{3.2711in}{2.4624in}{0pt}{\Qcb{Scaled density and
%pressure vs pressure/density}}{}{fig1.eps}%
%width 3.2711in;  height 2.4624in;  depth 0pt;  original-width 5.589in;
%original-height 4.1959in;  cropleft "0";  croptop "1";  cropright "1";
%cropbottom "0";  filename 'fig1.eps';file-properties "XNPEU";}}}%
%BeginExpansion
\begin{figure}
[h]
\begin{center}
\includegraphics[
height=2.4624in,
width=3.2711in
]%
{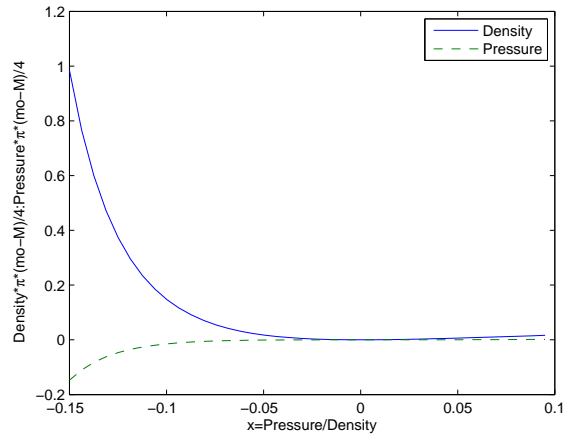}%
\caption{Scaled density and pressure vs pressure/density}%
\end{center}
\end{figure}
%EndExpansion

In general the equation of state does not have any of the usual simple
forms.\ It is a cubic equation in pressure and, depending on the value of
$\sigma,$ can contain simpler equations of state in the root structure.
Rewriting Eq.(\ref{p-square}), one finds%
\end{subequations}
\begin{align}
\overline{p}:=p(\pi/4)(m_{0}-M)  & \nonumber\\
\overline{\sigma}:=\sigma(\pi/4)(m_{0}-M)  & \nonumber\\
\overline{p}^{3}(64\overline{\sigma}-2)+\overline{p}^{2}(48\overline{\sigma
}^{2}-\overline{\sigma})+12\overline{\sigma}^{3}\overline{p}+\overline{\sigma
}^{4}  &  =0 \label{p-bar-cubed}%
\end{align}
The equation of state, Eq.(\ref{p-bar-cubed}), is plotted in Figure 2 showing
the regions where, for a single density value, there are three real roots, two
real roots, and one real root giving associated values of the pressure.\ The
negative pressure region corresponds to the layer tension.\ In this region,
for small density, there are two possible tensions for each value of density
while for larger density, there is a 1-1 non-linear relationship.
%TCIMACRO{\FRAME{fhFU}{3.2711in}{2.4624in}{0pt}{\Qcb{Scaled density vs scaled
%pressure for the equation of state}}{}{fig2.eps}%
%width 3.2711in;  height 2.4624in;  depth 0pt;  original-width 5.589in;
%original-height 4.1959in;  cropleft "0";  croptop "1";  cropright "1";
%cropbottom "0";  filename 'fig2.eps';file-properties "XNPEU";}}}%
%BeginExpansion
\begin{figure}
[h]
\begin{center}
\includegraphics[
height=2.4624in,
width=3.2711in
]%
{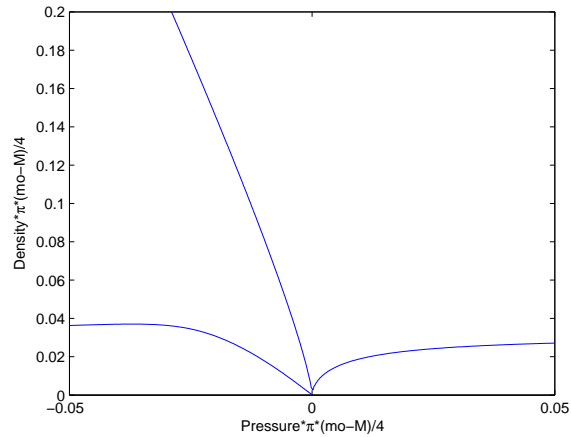}%
\caption{Scaled density vs scaled pressure for the equation of state}%
\end{center}
\end{figure}
%EndExpansion

Expanding Eq.(\ref{p-square}) one finds%
\[
P\approx\frac{4x^{3}}{\pi(m_{0}-M)}(1-10x+...)
\]
For $x<<1$ the equation of state appears approximately polytropic with index
$2$ \cite{Cha57}.\ Layers are most commonly described as perfect fluids or
scalar fields. Within the perfect fluid models, several EOS \cite{KMM06}%
,\cite{MHI02},\cite{Lob06} have been suggested to analyze the layer stress
energy content. \ A commonly discussed form is
\begin{equation}
P=-n\sigma
\end{equation}
which plays a crucial role in 3+1 homothetic solutions \cite{CT71}. The layer
is dropping on a Schwarzschld/Lemaitre geodesic. The actual equation of state
does not change but it can be modeled with simpler forms having time dependent
parameters. \ This is a different point of view than many dynamic layer
treatments which assume a static equation of state and let that drive the
layer dynamics.\ For example, Eq.(\ref{p-sigma}) for the Lemaitre layer
implies that $n(\tau)$ changes as the layer drops with the n value given by
\begin{equation}
n(\tau)=-x=\frac{\beta_{-}-1}{4\beta_{-}}.
\end{equation}
$n(\tau)$ runs through the range $0<n<1/4$ for the positive density layer. We
note that $x=0$ is the dust solution corresponding to $m_{0}=M$, the vanishing
layer condition, or to very large $R$.\ Positive $\beta_{-}>1$ describes an
inward moving layer with positive layer density, mass and tension. The
constraint $m_{0}>M$ implies that $r=M$ is inside the $r=2m_{0}$ trapped
surface.\ The physical picture of the layer dropping inward toward an interior
object of mass $M$ implies that the layer will cross the $2m_{0}$ horizon. The
value of $x$ at the horizon, $x_{H},$ is parameterized by the ratio $M/m_{0}$%
\begin{equation}
4x_{H}=\frac{1}{\sqrt{2-M/m_{0}}}-1.
\end{equation}
The maximum value of $x_{H}$, $x_{H}\sim-0.07,$ occurs for a Minkowski, $M=0$
interior, with $x_{H}$ decreasing as the size of the interior $M$ approaches
$m_{0}$. With $M=0$, the tension of the layer is $P=-.07\sigma$ at $R=2m_{0}$,
increasing inward to $-1/4$ at the $R=0$ singularity. The layer mass at
$R=2m_{0}$ is less than $m_{0}$, since $m_{0}$ will include the layer binding energy.

The region, $m_{0}<M,$ describes a layer with negative density and mass and
with $\beta_{-}<1,$ with $x$ positive. For this case $R=2m_{0}$ lies inside
$R=2M$. From Eq.(\ref{beta-minus}), $\beta_{-}=0$ at$\ R=2(M-m_{0})$.\ At this
radius, the layer has infinite tension.\ The interior Schwarzschild solution
could be the exterior of a third core fluid region with a boundary at $R>2M$,
eliminating trapped surfaces from the vacuum motion.

Another indication of the complexity of the Lemaitre layer is the sound
speed,
\[
V_{s}^{2}:=\frac{\partial P}{\partial\sigma}.
\]
Using the equation of state, this is%
\[
V_{s}^{2}=\frac{x}{2(1+x)}(3+8x+8x^{2}).
\]
$V_{s}^{2}$ is negative over the range $-1/4<x<0$ indicating instabilities in
the layer \cite{KT90}.

\section{Discussion}

In this paper we have discussed the geodesic collapse of an Israel layer in an
exterior Schwarzschild metric.\ The geodesic drop of the layer determines the
dynamic development of the layer's motion. Although their geodesic motion is
simple, Lemaitre layers are physically interesting structures because of their
complex equation of state, which can approximate some of the simpler equations
of state commonly used in layer dynamics. Some of the simplest equations of
state can be described by time dependent models of the actual stress density
relation.\ A barotropic equation of state is often used to describe Israel
layers, for example in effective potential treaments \cite{GV89}%
,\cite{Lak79},\cite{LW86} and discussion of bubble structures \cite{Ips87}.
\ In the simple model considered here, the layer can have a time dependent
barotropic relation, $P=-n(\tau)\sigma$. It has an $R\rightarrow0$ limiting
value $P=-\sigma/4$, with $n(\tau)$ still very small as the layer reaches the
$R=2m_{0}$ surface.

The range of $P/\sigma$ for positive density, has a lower bound of
$-1/4$.\ This is a new restriction on the range of the stress/density ratio
for collapsing layers.\ Garfinkle and Vuille \cite{GV89} considered the layer
between two Schwarzschild metrics with no motion restrictions. \ The value
$P/\sigma=-1/2$ was identified as a boundary for their effective potential
\cite{Lak79}, \cite{LW86} to have stationary points.\ The effective potential
for the Lemaitre layer is simple with no stationary points or turning points
and the boundary $P=-\sigma/4$ occurs at the point $R=0$. \ 

Fluids collapsing along geodesics in a Schwarzschild metric have been
considered by Lynden-Bell and Lemos \cite{L-BL88}, generalizing the
description of a Newtonian cold gas collapsing from rest at infinity
\cite{Pen69}. The collapse of the Lemaitre layer and the cold gas have some
similarities. They are both described by the the same $R(\tau)$ and for either
case Eq.(\ref{cap-R}) can be written as
\[
\frac{4m_{0}}{3}[\frac{R_{i}(\tau)}{2m_{0}}]^{3/2}=\tau_{c}-\tau_{i}%
\]
where $\tau_{i}$ is the proper time to reach $R_{i}$ and $\tau_{c}$ is the
time to reach $R=0$.\ In the less limiting case of a Newtonian sphere with
interior mass $m$, Lynden-Bell and Lemos show that $\tau_{c}$ can be written
as a function of the mass%
\[
\tau_{c}=Bm^{b}+\tau_{0},
\]
where $\tau_{0}$ is a constant of integration.\ Generalizing this relation to
Einstein's gravity, $b=1$ is identified as a value of special interest
corresponding to metric homothetic symmetry.\ For the Lemaitre layer,
$\tau_{c}$ is just the constant $c_{1}=\tau_{0}$, corresponding to $B=0$ in
the cold gas solution. $\tau_{c}$ is related to the time $\tau_{H}$ for a
layer to reach the $R=2m_{0}$ surface by
\[
\tau_{c}=4m_{0}/3+\tau_{H}.
\]
The Lemaitre layer can be considered as a special member of the $b=1$, $B=0$
geodesic solutions which include cold gas collapse. One also notes that the
exterior bounding Lemaitre metric exhibits second kind kinematic similarity
\cite{CH89}, reflected by $R$($\tau)$ in Eq.(\ref{cap-R}), which is
transferred to the layer and maintained thoughout the motion.

There is growing interest in Israel layers as objects of research rather than
an aid in matching metrics. \ The increasing interest in expanding the
physical properties of Israel layers has led to descriptions incorporating
layer parameters like shell thickness \cite{KM02}, \cite{KKM06} and surface
packing fraction \cite{Kri05}. While Israel layers are still linked to the
properties of the two bounding spaces, even simple layers such as the Lemaitre
layer considered here, require more physical content than the density and
pressure coming from simple perfect fluid or polytropic models. \ The time
dependent models for the actual stress density relations described here,
represent one way these simple models can still be used effectively and
related to the literature using simple equations of state.

\appendix{}

\section{Extrinsic curvatures}

The stress-energy of the layer is determined by jumps in the extrinsic
curvatures, $K_{ab}$ of the bounding metrics across the Israel layer. The
bounding metrics considered here are%
\begin{equation}
ds^{2}=-f_{\pm}dt^{2}+(1/f_{\pm})dr^{2}+r^{2}d\Omega^{2}%
\end{equation}
with ($f_{+}$ $,$ $f_{-}$) = ($1-2m_{0}/r,$ $1-2M/r$). The layer is described
by $r=R(\tau)$ and $t=T_{\pm}(\tau)$. The velocity of the layer as seen by
observers moving with the layer in the bounding spaces is $U_{\pm}^{i}%
=(\dot{T}_{\pm},\dot{R},0,0)$, with corresponding outward pointing normal
vector $n_{a\pm}=(-\dot{R},\overset{\cdot}{T}_{\pm},0,0)\ $\cite{Poi04}%
$.$\ The velocity normalization imposes%
\begin{equation}
f_{\pm}(R)\dot{T}_{\pm}^{2}-\frac{\dot{R}^{2}}{f_{\pm}(R)}=1.
\end{equation}
The normalization is used to define a function, $\beta_{\pm}$%
\begin{equation}
\beta_{\pm}^{2}:=f_{\pm}^{2}(R)\dot{T}_{\pm}^{2}=f_{\pm}(R)+\dot{R}^{2}.
\end{equation}
We choose $\beta_{\pm}>0.$\ $\beta_{\pm}<0$ corresponds to $\tau
\rightarrow-\tau.$\ The metric of the layer from both sides of the boundary
is
\[
ds^{2}=-d\tau^{2}+R^{2}(\tau)d\Omega^{2}.
\]
The extrinsic curvature is defined in terms of the normal to the layer,
$n_{a}$ and the projection operator onto the layer, $h_{ab}$ as%
\begin{equation}
K_{ij}:=n_{a;b}h_{i}^{a}h_{j}^{b}%
\end{equation}
The two extrinsic curvatures are $K_{\theta}^{\theta}$ and $K_{\tau}^{\tau}.$
$K_{\ \theta}^{\theta}$ is easily calculated from the definition and one finds%
\begin{equation}
K_{\theta\pm}^{\theta}=\frac{\beta_{-}}{R}=\frac{\sqrt{f_{\pm}(R)+\dot{R}^{2}%
}}{R}.
\end{equation}
where we use $\beta_{-}$ to indicate the range of $\beta_{\pm}$ to be used.
Using the infalling geodesic value for $\dot{R}=-\sqrt{2m_{0}/R}$ we have%
\begin{equation}
K_{\theta\pm}^{\theta}=\frac{\beta_{-}}{R}=\frac{\sqrt{f_{\pm}(R)+2m_{0}/R}%
}{R}.
\end{equation}
$K_{\tau}^{\tau}$ is more easily calculated from the acceleration $K_{\tau
}^{\tau}=n^{i}a_{i}.$ The acceleration vector is%
\begin{equation}
a^{i}=[\ddot{T}+\frac{\dot{T}\dot{f}}{f},\text{ }\ddot{R}+\frac{\dot{f}}%
{2\dot{R}},\text{ }0,\text{ }0].
\end{equation}
Dotting with the normal and using the definition of $\beta_{\pm}$, one finds
\begin{equation}
K_{\tau\pm}^{\tau}=\frac{\dot{\beta}_{\pm}}{\dot{R}}\ .
\end{equation}
All cases will have the same exterior functions. In the interior, the
extrinsic curvatures are%
\begin{align}
\beta_{-}  &  =\sqrt{1+\frac{2(m_{0}-M)}{R}}\\
K_{\theta-}^{\theta}  &  =\beta_{-}/R\\
K_{\tau-}^{\tau}  &  =\frac{M-m_{0}}{R^{2}\beta_{-}}.
\end{align}


\begin{thebibliography}{99}                                                                                               %


\bibitem {Isr66}W. Israel, Nuov. Cim. \textbf{44B}, 1 (1966).

\bibitem {Isr67}W. Israel, Nuov. Cim. \textbf{48B}, 463\ (1967).

\bibitem {Bar91}C. Barrabes and W. Israel, Phys. Rev. D \textbf{43}, 1129 (1991).

\bibitem {Poi04}E. Poisson, \textit{A Relativist's Toolkit} (Cambridge
University Press, Cambridge, England, 2004) p.93.

\bibitem {CI68}V. de la Cruz and W. Israel, Phys. Rev. \textbf{170}, 1187 (1968).

\bibitem {ML97}P. Musgrave and K. Lake, Classical Quantum Gravity \textbf{14},
1285 (1997).

\bibitem {BH98}C. Barrabes and P. A. Hogan, Phys. Rev. D \textbf{58}, 044013 (1998).

\bibitem {BKT87}V.A. Berezin, V.A. Kuzmin and I.I. Trachev, Phys. Rev. D
\textbf{36}, 2919 (1987).

\bibitem {Ans02}S. Ansoldi, Classical Quantum Gravity \textbf{19}, 6321 (2002).

\bibitem {Ans07}S. Ansoldi, arXiv:gr-qc/0701082. \textbf{(}\emph{To appear in
the proceedings of the Eleventh Marcel Grossmann Meeting on General
Relativity, July 23-29, 2006, Freie Universitaet Berlin, Berlin, GR}

\bibitem {OR07}L. Ortiz and M.P. Ryan, Jr, Gen. Relativ. Gravit. \textbf{39},
1087 (2007).

\bibitem {AJ05}M.\ Ambrus and P. Jahicek, Phys. Rev. D \textbf{72, }064025 (2005).

\bibitem {Kri05}J.P. Krisch, J. Math. Phys. (N.Y.) \textbf{46}, 042506 (2005).

\bibitem {BGK06}S.Bayin, E.N. Glass and J.P. Krisch, J. Math. Phys. (N.Y.)
\textbf{47}, 012501 (2006).

\bibitem {Kri06}J.P. Krisch, J. Math. Phys. (N.Y.) \textbf{47}, 122501 (2006).

\bibitem {GW07}E. Gravanis and S. Willison, Phys. Rev. D \textbf{75,
}084025\textbf{\ }(2007).

\bibitem {RKS06}F. Rahaman, M. Kalam and S. Chakraborti, arXiv:gr-qc/0611134.
\emph{(Int. J. Mod. Phys. D, to be published)}

\bibitem {ES05}E.F. Eiroa and C. Simeone, Phys. Rev. D \textbf{71, }127501 (2005).

\bibitem {BI05}C. Barrabes and W. Israel, Phys. Rev. D \textbf{71, }064008 (2005).

\bibitem {LC05}F.S.N. Lobo and P. Crawford, Classical Quantum Gravity
\textbf{22},.4869 (2005).

\bibitem {MS93}M. Mars and\ J.M.M.\ Senovilla, Classical Quantum Gravity
\textbf{10}, 1865 (1993).

\bibitem {Ver02}R. Vera, Classical Quantum Gravity \textbf{19}, 5249 (2002).

\bibitem {Mar05}M. Mars,\ Classical Quantum Gravity \textbf{22}, 3325 (2005).

\bibitem {MMV07}M. Mars, F. C. Mena and R. Vera, arXiv/gr-qc/07040078.

\bibitem {KMM06}J. Kijowski, G. Magli and D. Malafarina, Gen. Relativ. Gravit.
\textbf{38}, 1697 (2006).

\bibitem {GV89}D. Garfinkle and C. Vuille, Classical Quantum Gravity
\textbf{6}, 1819 (1989).

\bibitem {LL07}P.D. Lasky and A.W.C. Lun, Phys. Rev. D \textbf{75}, 104010 (2007).

\bibitem {MM01}P.O. Mazur and E. Mottola, arXiv:gr-qc/0109035.

\bibitem {MM04}P.O. Mazur and E. Mottola, Proc. Nat. Acad. Sci. \textbf{111},
9545 (2004).

\bibitem {VW04}M. Visser and D.L. Wiltshsire, Classical Quantum Gravity
\textbf{21}, 1135 (2004).

\bibitem {Car05}B. Carter, Classical Quantum Gravity \textbf{22}, 4551 (2005).

\bibitem {CFV05}C. Cattoen, T. Faber and M. Visser, Classical Quantum Gravity
\textbf{22}, 4189 (2005).

\bibitem {LA06}F.S.N. Lobo and A.V.B. Arellano, Classical Quantum
Gravity\textbf{\ 24, }1069 (2007).

\bibitem {deBHI+06}A. deBenedictis, D. Horvat, S. Ilijic, S. Kloster and
K.S.Viswanathan, Classical Quantum Gravity \textbf{23}, 2303 (2006).

\bibitem {BN07}A. E. Broderick and R. Narayan, Classical Quantum Gravity
\textbf{24, }659\textbf{\ }(2007).

\bibitem {KM02}S. Khakshournia and R. Mansouri, Gen. Relativ. Gravit.
\textbf{34}, 1847 (2002).

\bibitem {KKM06}Sh. Khosravi, S. Khakshournia and R. Mansouri, Classical
Quantum Gravity \textbf{23, }5927\textbf{\ }(2006). 

\bibitem {MHI02}H. Maeda, T. Harada and H. Iguchi, Phys. Rev.D \textbf{66},
27501 (2002).

\bibitem {CY90}B.J. Carr and A. Yahil, Astrophys. J. \textbf{360}, 330 (1990).

\bibitem {Vis95}M. Visser, \textit{Lorentzian wormholes: From Einstein to
Hawking}, (AIP Press, New York, 1995) p.183

\bibitem {Cha57}S. Chandrasekhar, \textit{An Introduction to the Study of
Stellar Structure}, (Dover, New York, 1957), Ch. 6.

\bibitem {Lob06}F.S.N. Lobo, Phys. Rev. D\textbf{\ 75, }024023 (2007).

\bibitem {CT71}M.E. Cahill and A.H. Taub, Comm. Math. Phys. \textbf{21}, 1 (1971).

\bibitem {KT90}E.W. Kolb and M.S. Turner, \textit{The Early Universe},
(Addison-Wesley, MA, 1990), p. 342.

\bibitem {Lak79}K. Lake, Phys. Rev. D \textbf{19, }2847 (1979).

\bibitem {LW86}K. Lake and R. Wevrick, Canadian Journal of Physics
\textbf{64}, 165 (1986).

\bibitem {Ips87}J.R. Ipser, Phys. Rev. D \textbf{36}, 1933 (1987).

\bibitem {L-BL88}D. Lynden-Bell and J.P.S. Lemos, MNRAS \textbf{233}, 197 (1988).

\bibitem {Pen69}M.V. Penston, MNRAS \textbf{144}, 425 (1969).

\bibitem {CH89}B. Carter and R.N. Henriksen, Ann. Physique Supp. \textbf{14},
47 (1989).
\end{thebibliography}
\end{document}